\documentclass{emulateapj}
\usepackage{epstopdf}
\usepackage{hyperref}
\usepackage{verbatim}
\usepackage{hyperref}

\newcommand{\kepler}{{\it Kepler }}

\newcommand{\kp}{$K_P$}

\slugcomment{Accepted to ApJ}

\shorttitle{Parameters of the Coolest {\it Kepler} Targets}
\shortauthors{Mann et al.}

\begin{document}

\title{They might be giants:\\ luminosity class, planet occurrence, and
  planet-metallicity relation of the coolest {\it Kepler} target stars}
  
 \author{Andrew W. Mann\altaffilmark{1}, Eric Gaidos\altaffilmark{2}, S\'{e}bastien L\'{e}pine\altaffilmark{3}, Eric J. Hilton\altaffilmark{2}}
  
\altaffiltext{1}{Institute for Astronomy, University of Hawai'i, 2680 Woodlawn Dr, Honolulu, HI 96822} 
\altaffiltext{2}{Department of Geology \& Geophysics, University of Hawai'i, 1680 East-West Road, Honolulu, HI 96822} 
\altaffiltext{3}{Department of Astrophysics, American Museum of Natural History, New York, NY 10024}

\begin{abstract}
We estimate the stellar parameters of late K and early M type \kepler target stars. We obtain medium resolution visible spectra of 382 stars with $K_P-J>2$ ($\simeq$ K5 and later spectral type). We determine luminosity class by comparing the strength of gravity-sensitive indices (CaH, K I, Ca II, and Na I) to their strength in a sample of stars of known luminosity class. We find that giants constitute $96\pm1\%$ of the bright ($K_P< 14$) \kepler target stars, and $7\pm3\%$ of dim ($K_P > 14$) stars, significantly higher than fractions based on the stellar parameters quoted in the \kepler Input Catalog (KIC). The KIC effective temperatures are systematically ($110^{+15}_{-35}$~K) higher than temperatures we determine from fitting our spectra to PHOENIX stellar models. Through Monte Carlo simulations of the \kepler exoplanet candidate population, we find a planet occurrence of $0.36\pm0.08$ when giant stars are properly removed, somewhat higher than when a KIC log~$g>4$ criterion is used ($0.27\pm0.05$). Lastly, we show that there is no significant difference in $g-r$ color (a probe of metallicity) between late-type \kepler stars with transiting Earth-to-Neptune sized exoplanet candidates and dwarf stars with no detected transits. We show that a previous claimed offset between these two populations is most likely an artifact of including a large number of misidentified giants.
\end{abstract}


\section{Introduction}\label{sec:intro}
The NASA \kepler mission \citep{Borucki:2010lr} has ushered exoplanet science into a new phase of analysis based on the statistics of large samples.  Among the more elementary statistics derived from {\it Kepler} results are the planet occurrence around stars \citep[][henceforth H11]{Howard:2011ul}, the distribution of planet size \citep[or mass][]{Wolfgang:2011uq,Gaidos:2012lr}, correlations between the presence of planets and the properties of the host stars \citep[e.g.][henceforth SL11]{Schlaufman:2011pd}, and the characteristics of multi-planet systems \citep{Fabrycky:2012lr}. These findings yield important constraints on models of planet formation and evolution, and are best established for solar-type stars (late F through early K spectral types) because they constitute the vast majority of {\it Kepler} targets. 

The results of {\it Kepler} were first preceded by the findings of radial velocity surveys of solar-type stars. More than 15\% of dwarf stars have close-in ($\sim$0.25~AU) planets with orbital periods less than 50 days \citep{2010Sci...330..653H, Howard:2011ul} and this fraction increases with orbital period \citep{Mayor:2011fj}. The same authors find that planet occurrence is inversely related to planet mass or radius, with ``super-Earths'' outnumbering Jupiter-size planets by more than an order of magnitude.  Around solar-type stars, the presence of giant planets is strongly correlated with super-solar metallicity \citep{1997MNRAS.285..403G, 2004A&A...415.1153S, Fischer:2005yq, Johnson:2010lr}, but this correlation does not appear to hold for smaller planets \citep{2008A&A...487..373S,2009A&A...496..527B,Mayor:2011fj}. As with results from {\it Kepler}, these findings are primarily for solar-type stars because many nearby representatives are bright enough for ground-based Doppler radial velocity observations. 

Very cool (late K and early M type) dwarf stars have become popular targets of planet searches \citep[e.g.][]{Charbonneau:2009rt, Vogt:2010fr, Bean:2010ys, Mann:2011qy, Apps:2010zr, Fischer:2012lr}. Planets around cool stars are easier to detect because of the stars' smaller masses and radii. Furthermore, because these stars are less luminous, close-in and thus detectable planets can still orbit within the ``habitable zone,'' where an Earth-like planet would avoid the ``snowball'' or runaway greenhouse climate states \citep{Gaidos:2007ly}.  However, the statistics of planets around these stars are poorly established. These stars are underrepresented in magnitude-limited Doppler surveys as well as the \kepler target list.  Only 2\% of \kepler target stars are classified as possible M types (cooler than 4000~K), whereas $>$70\% of all stars within 20~pc are M dwarfs \citep{Henry:1994fk, 2003PASP..115..763C,Reid:2004lr}.  

Nevertheless, {\it Kepler} data has been used to draw two important conclusions about late-type exoplanet hosts. First, H11 found that the frequency of stars with planets on close-in (P$<50$d) orbits rises with decreasing effective temperature through early K-type and that an even higher fraction of M dwarf stars may host such planets. Second, SL11 claimed that late K dwarf stars, but not solar-type stars, hosting super-Earth to Neptune sized candidate transiting planets are more metal rich than stars for which transits have not been detected. These findings offer potential tests of theories of planet formation \citep{Fischer:2005yq, 2008ApJ...673..502K, Cumming:2008bh}

\kepler targets are selected from the \kepler input catalog (KIC) based on the ability of the mission to find transiting planets, especially in the habitable zone; ideally, the target catalog should consist exclusively of dwarf stars for which the signal of a transiting planet is largest, and exclude sub-giant and giant stars.  \citet{Brown:2011fj} used D51 (Mg Ib line) photometry and Sloan $g$-D51 color to exclude giants, however this is also sensitive to temperature and metallicity and is not available for all targets.  The KIC includes Sloan ($griz$) and 2MASS ($JHK$) magnitudes; stellar parameters are estimated by forward modeling of the photometric data with the synthetic spectra of \citet{Castelli:2004lr}, and effective temperature $T_{eff}$, gravity log~$g$, and metallicity [M/H] as free parameters.  Stellar mass and distance are then estimated using luminosity, $T_{eff}$, and log~$g$ from the stellar evolutionary models of \citet{2000A&AS..141..371G}.  The combination of stellar mass and log~$g$ then yields a stellar radius.

\citet{Brown:2011fj} state that KIC radius estimates have average errors of 35\% and are not reliable for stars cooler than 4000~K. H11 point out that, because of the difficulty in constraining log~$g$, the radii of some stars, particularly sub-giants, may be underestimated by a factor of 2 or more in the KIC.  \cite{Gaidos:2012lr} found that consistency between the {\it Kepler} candidate planet catalog and the M2K Doppler survey could be achieved if the former was incomplete compared to estimates based on KIC radii. They further point out that \kepler planet candidates were conspicuously sparse among late K stars with colors that are shared by both dwarfs and giant stars.  Finally, \citet{Muirhead:2012pd} (henceforth M11) showe that KIC estimates for the radii of many \kepler M dwarfs hosting planets are smaller than KIC values by as much as a factor of two. This discrepancy is not to be confused with the 5-10\% radius difference between radii of the most refined models and measurements by interferometry and observations of eclipsing binaries \citep[e.g.][]{2010A&A...514A..97L, Kraus:2011rr}.

Reliable stellar parameters are a prerequisite for robust statistical analysis of planets, especially transiting planets. These are needed not only for stars for which planet candidates have been detected (referred to as \kepler Objects of Interest or KOIs), but also for the target sample as a whole. The radius of a planet producing a given transit depth is proportional to the radius of its host star. Likewise, the transit signal produced by a planet of a given radius - and hence its detectability around a star in the survey - also depends on stellar radius. If some target stars are actually larger or even giant stars, then planets are less likely to be detected in that sample, which means that the most likely occurrence rate of those planets is higher. For M dwarf stars in general, and particularly for the coolest \kepler target stars, parameters such as radius are uncertain or even very unreliable \citep[e.g.][M11]{Johnson:2012fk}.

\citet{Brown:2011fj} metallicities are reliable to 0.4~dex for solar-type stars, but are essentially useless for stars with T$_{eff}<4000$. Instead, SL11 use Sloan $g-r$ colors for a given $J-H$ range (a proxy for spectral type) as an indicator of the amount of Fe line blanketing at blue wavelengths, and hence metallicity. They construct mean $g-r$ vs. $J-H$ loci for KOIs and \kepler stars without identified transits. They find a significant difference between the $g-r$ colors of the two populations for stars with $J-H \approx 0.62$, corresponding to late K-type stars. Based on stellar models, SL11 argue that the late-type KOIs are $\simeq0.2$~dex more metal rich than {\it Kepler} targets with no detected transit. However, K giants are significantly bluer than dwarfs in $g-r$, for the same $J-H$ \citep{Yanny:2009mz}. Thus, contamination of the \kepler target sample by giants would shift the locus of target stars to bluer $g-r$, but would not affect the KOI locus, as planets are less detectable, or completely undetectable around giant stars.  Realizing this, SL11 constructed and analyzed artificial mixed data sets to estimate that a 10-30\% contamination by giants would also produce the observed offset.

M11 use the equivalent widths of atomic lines in the $K$ (2.2 $\mu$m) band \citep{Rojas-Ayala:2012uq} and their measurements of late-type KOIs' metallicities are consistent with, or slightly metal-poor (median [M/H] = \,-0.10) compared to the solar neighborhood \citep[M/H $\simeq -0.05$][]{Johnson:2009fk}. SL11 and M11 are consistent with each other if the {\it Kepler} target list itself is biased toward metal-poor M dwarfs, or if the offset found by SL11 is due to high giant contamination in {\it Kepler} late-type target stars. 

Moderate resolution spectra are nearly always sufficient to distinguish K and M giants from their dwarf cousins.  In addition \citet{Ciardi:2011lr} showed that some giant stars can be identified based on $JHK$ photometry alone. In this paper, we combine moderate resolution spectra of a sample of Kepler targets with KIC photometry to refine the planet occurrence rate for late-type stars calculated by H11, and determine if the giant fraction is high enough to explain the color offset observed by SL11. In Section \ref{sec:obs} we present spectroscopy of a representative sample of late-type \kepler target stars. In Section \ref{sec:class} we use both spectroscopy and photometry to derive luminosity classes and calculate the giant fraction for late-type {\it Kepler} target stars. In Section \ref{sec:occurrence} we use this information, plus radii based on stellar evolutionary models, to refine the planet occurrence around these stars. In Section \ref{sec:metallicity} we calculate and compare the mean $g-r$ colors (as metallicity proxies) of KOIs and a {\it bona fide} dwarf sample, and show how and why our results differ from those of SL11.

\section{Sample, Observations, and Reduction} \label{sec:obs} Because derived KIC parameters may not always be reliable, we instead select our sample using photometry. A sample of stars with $V-J>2.5$ will include $>98\%$ of all M dwarfs, as well as most of the K7 dwarfs in the sample \citep[][henceforth LG11]{Lepine:2011vn}. Although 2MASS $J$ magnitudes are available for almost the entire sample, $V$ magnitudes are not. {\it Kepler} magnitudes ($K_P$), however, are available for all target stars. For M0 stars, $K_P-V\simeq-0.43$\footnote{keplergo.arc.nasa.gov/CalibrationZeropoint.shtml} so we conservatively select stars with $Kp-J>2$ observed in Quarters 0-2 by {\it Kepler} and retrieved from the Multimission Archive (STScI). We remove stars with a contaminating star within 1 arc second. 

\begin{figure}
\centering
\includegraphics[width=8cm]{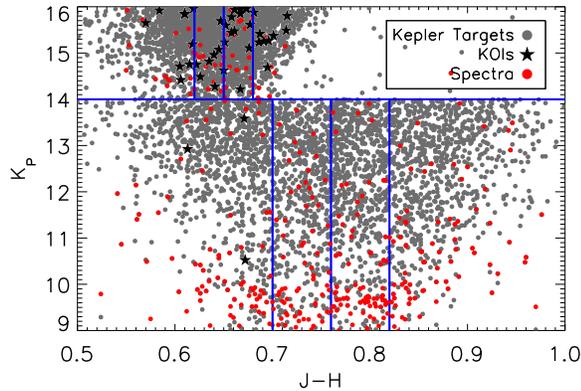}
\caption{\kepler magnitude vs. $J-H$ color for Quarter 0-2 \kepler target stars with $K_P-J>2$ (grey circles), KOIs (black stars), and targets with spectra from this program (red circles). Our observing bins (see Section \ref{sec:obs}) are marked by blue lines. There is a clear difference between the colors of bright ($K_P<14$) and dim ($K_P>14$) {\it Kepler} target stars, resulting in a very different distribution of colors. The great majority of KOIs are faint, and have bluer $J-H$ colors. For this reason we divide the sample into $J-H$ bins, and treat bright ($K_P<14$) and dim ($K_P>14$) \kepler target stars as two independent samples. \label{fig:JHkp}}
\end{figure}

\begin{figure}
\centering
\includegraphics[width=8cm]{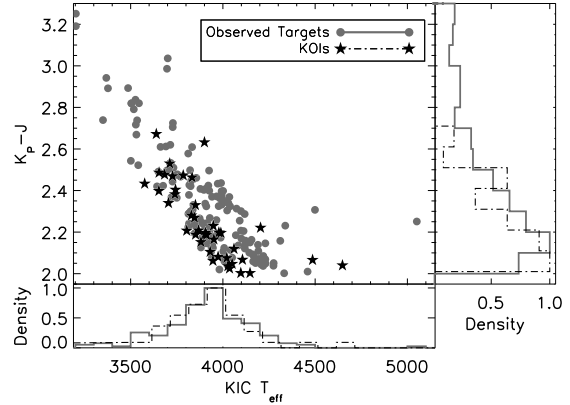}
\caption{Distribution of KIC effective temperatures and $K_P-J$ colors for target stars (grey circles and grey solid histogram) and KOIs (black stars and black dashed histogram). The bulk of the stars in our spectroscopic sample are M dwarfs ($T_{eff}<4000~K$) if we assume KIC $T_{eff}$ values are accurate. Histograms for KOIs and observe targets offset slightly from each other for clarity (although the bins for each sample are the same). Note that not all stars have effective temperatures listed in the KIC; points lacking $T_{eff}$ values are not shown in the center plot or bottom histogram, but are included in the $K_P-J$ histogram. \label{fig:sample}}
\end{figure}

Bright \kepler target stars were selected in a fundamentally different way from dim stars \cite[see Figure~\ref{fig:JHkp} and ][]{Batalha:2010fk}. We separately analyzed dim ($K_P>14$) and bright ($K_P<14$) stars. \citet{Bessell:1988qy} showed that giant stars tend to have more extreme $J-H$ colors than their dwarf counterparts. However, we wanted to investigate how {\it misidentified} giant stars in the KIC are distributed with $J-H$ color. Thus we further subdivided our sample into four $J-H$ color bins: $J-H\le0.70$, $0.70<J-H\le0.76$, $0.76<J-H\le0.82$, and $0.82<J-H$ for the bright stars and $J-H\le0.62$, $0.62<J-H\le0.65$, $0.65<J-H\le0.68$, and $J-H>0.68$ for the dim stars. Color bins were designed such that each contains a similar number of stars. We observed a sample of stars within each bin, selected randomly with respect to $J-H$. We observed more bright stars because they are more observationally accessible, although we observed targets spanning all {\it Kepler} magnitudes to detect trends with $K_P$. In total we observed 382 stars covering $6.5 <$ \kp $ < 16$,  $0.40 < J-H < 1.00$ and KIC effective temperatures $3200 < $ $T_{eff}$$ < 5050$~K. The distribution of observed targets is shown in $J-H$ and KIC $T_{eff}$ space in Figure~\ref{fig:sample}. A list of observed targets is given in Table \ref{tab:observations}.

\begin{table*} 
\caption{Parameters of Observed {\it Kepler} Targets$^a$}
 \centering
\begin{tabular}{l l l l l l l l l l l l}
\hline \hline
\multicolumn{4}{c}{KIC Parameters} & \multicolumn{4}{c}{Derived from Spectra}\\
KIC ID & $K_P$ & log~$g$ & T$_{eff}$ [K] & Instrument$^b$ & Luminosity Class$^c$ & T$_{eff}$ [K] & $\sigma_T$ [K] \\  \hline
   3001835$^d$ & 13.5 & --- & --- & SNIFS & Giant & 3800 &  60 \\ 
   8881126 & 15.8 &  4.6 & 3890 & SNIFS & Dwarf & 3720 &  60 \\ 
  10717091$^d$ & 10.3 & --- & --- & MkIII & Giant & 3790 &  60 \\ 
  10406398$^d$ & 15.9 & --- & --- & SNIFS & Giant & 3310 & 100 \\ 
   8426324 & 10.3 &  2.2 & 3526 & CCDS & Giant & 3480 &  90 \\ 
   3455941$^d$ & 10.9 & --- & --- & CCDS & Giant & 4260 &  40 \\ 
   6032907 & 15.3 &  4.5 & 3938 & SNIFS & Dwarf & 3830 &  70 \\ 
  10064712$^d$ &  9.4 & --- & --- & CCDS & Giant & 4050 &  60 \\ 
   5112438$^d$ & 10.7 & --- & --- & MkIII & Giant & 3600 &  70 \\ 
   5732026$^d$ &  9.5 & --- & --- & MkIII & Giant & 3860 &  70 \\ 
  10593779$^d$ &  8.9 & --- & --- & CCDS & Giant & 3930 &  60 \\ 
   4818175$^d$ &  7.6 & --- & --- & CCDS & Giant & 3930 &  50 \\ 
   9175009 & 14.2 &  4.6 & 3836 & SNIFS & Dwarf & 3870 &  70 \\ 
  12417370 &  9.0 &  2.5 & 4098 & CCDS & Giant & 3900 &  60 \\ 
  10843322 & 15.0 &  4.4 & 3650 & SNIFS & Dwarf & 3580 &  80 \\ 
   6695442$^d$ & 13.4 & --- & --- & SNIFS & Giant & 3630 &  70 \\
   ... & & & & \\
   \hline
\end{tabular}
\label{tab:observations}
\tablecomments{$^a$ The complete table is available in the electronic version of Astrophysical Journal and at \url{http://ifa.hawaii.edu/~amann/Table1_full.txt}.This table is provided for guidance on the fields and format. \\
$^b$SNIFS = SuperNova Integral Field Spectrograph, CCDS = Boller \& Chivens CCD Spectrograph, MkIII = Mark III spectrograph. SNIFS is attached to the University of Hawaii 2.2-meter telescope, and both CCDS and MkIII at the MDM Observatory 1.3m McGraw-Hill Telescope.\\
$^c$ Luminosity classes derived based on our spectra.\\
$^d$ No temperatures or log~$g$ values present in the KIC
}
\end{table*}

Observations were obtained between June 16 and Aug 28 (2011) with the SuperNova Integral Field Spectrograph \citep[SNIFS,][]{Lantz:2004fk} at the University of Hawaii 2.2m telescope on Mauna Kea and the Boller and Chivens CCD Spectrograph (CCDS) or the Mark III spectrograph (MkIII) at the MDM Observatory 1.3m McGraw-Hill telescope on Kitt Peak. SNIFS is an optical integral field spectrograph with $R \simeq 1300$ that splits the signal with a dichroic mirror into blue ($3000 - 5200$~\AA) and red ($5000 - 9500$~\AA) channels. SNIFS images were resampled with microlens arrays, dispersed with grisms, and focused onto blue- and red-sensitive CCDs.  Processing of SNIFS data was performed with the SNIFS pipeline, described in detail by \citet{Aldering:2006} and \citet{2010AIPC.1241..259P}. SNIFS processing included dark, bias, and flat-field corrections, assembling the data into red and blue 3D data cubes, and cleaning them for cosmic rays and bad pixels. After sky subtraction, the spectra are extracted with a PSF model, and wavelengths were calibrated with arc lamp exposures taken at the same telescope pointing as the science data. 

The CCDS and MkIII spectrographs cover $5700 - 9300$\AA\ and $4400 - 8300$\AA\ with $R\simeq1150$ and $\simeq2300$, respectively. Standard reduction of data taken with the CCDS and MkIII was performed with IRAF, following the practice of overscan subtraction, division by flat field, and extraction of the spectra. Spectra were wavelength-calibrated against NeArXe comparison arcs. All observations (including SNIFS) were flux-calibrated and telluric lines were removed based on observations of the NOAO primary spectrophotometric standards Feige 66, Feige 110, and BD+284211. All spectra had a median S/N of $>30$ in the 6000-7000\AA\ range, and the median S/N of all spectra in this range was 50.

Our spectroscopic set only covers $K_P-J>2.0$, but we also consider a separate `photometric sample' that includes stars with $0.56 < J-H < 0.66 \bigcup K_P-J > 2$. This is done so we can ensure coverage of the sample of late K stars used by SL11 (see Section~\ref{sec:metallicity}). The KIC includes $JHK$ photometry from 2MASS \citep{Skrutskie:2006lr} and visible-wavelength photometry through SDSS $griz$ and $D51$ filters. We add photometry from the Wide-field Infrared Survey Explorer \citep[WISE,][]{Wright:2010fk}, which includes 3.4$\mu$m, 4.6$\mu$m, 12$\mu$m, and 22$\mu$m bands.

\section{Luminosity Class} \label{sec:class}
We determine luminosity class by comparing the spectral indices or colors of \kepler target stars to those of stars drawn from `training sets' of known giants or dwarfs. We first discuss how we construct our training sets. We then explain our choice of indices and color-color relations, based on previous work on giant/dwarf discrimination and derived empirically from examination of the differences between the dwarf and giant training set. We use the colors and spectroscopic indices of stars in the training sets to construct a likelihood estimator, such that we can calculate the likelihood that a given star is a giant (or dwarf). That calculation is explained in Section~\ref{sec:apptokep}. 

\subsection{Training Sets}\label{sec:training}
We construct an uncontaminated set of dwarf stars from a sample of high proper motion-selected late dK and dM stars (LG11). The brightest ($J<9$) northern stars in the LG11 catalog have visible-wavelength spectra (L\'{e}pine et al. in prep), obtained with the same instruments and reduced in the same way as was done for \kepler targets observed for this paper. Although the sample from L\'{e}pine et al. (in prep) includes more than 1500 spectra, we construct our dwarf sample only from the 620 targets with spectra from SNIFS/UH2.2m, which includes the Ca II triplet feature at $8484-8662$\AA.

LG11 use $J-H$, and $H-K$ colors, combined with proper motion from SUPERBLINK \citep{2005AJ....129.1483L} and (for some targets) parallax information from Hipparcos \citep{van-Leeuwen:2005kx,van-Leeuwen:2007yq} to remove giant stars. Based on those stars in LG11 with parallaxes, we estimate that fewer than 0.5\% of the resulting sample will be giants. However, because of strict cuts in $J-H$ and $H-K$, the LG11 sample is incomplete and biased against dwarfs with much redder or bluer colors. LG11 also use a color cut of $V-J>2.7$ to select mostly M dwarfs. This excludes some mid- to late-K stars which will be included in our ($K_P-J>2 \bigcup 0.58<J-H<0.66$) color cut for the photometric sample (see Section~\ref{sec:obs}). We therefore add 60 late K and early M dwarfs included in the {\it Hipparcos} catalog that have UH2.2m spectra but lie outside the cuts imposed by LG11. These stars are confirmed to be dwarfs by their {\it Hipparcos} parallaxes. We also add 150 M dwarfs with spectra from SDSS, including 50 dwarf from \citet{West:2011fj}, with $r-J$ and $J-H$ colors consistent with our targets of interest. We verify that these targets are dwarfs using a cut with reduced proper motion, where the reduced proper motion in the SDSS $g$ band is:
\begin{equation}\label{eqn:RPM}
H_g = g + 5\log \mu + 5, 
\end{equation}
and $\mu$ is the proper motion in arcsec yr$^{-1}$. This quantity is similar to the absolute magnitude, such that giant stars will have much lower reduced proper motions than dwarfs of the same color. We only select SDSS stars with $H_g > 2.2(g - r) + 7.0$, and $\mu>15$ arcsec yr$^{-1}$, which we determine empirically from our UH2.2m targets with SDSS photometry. 

Our sample of $>$300 giant spectra is constructed from multiple catalogs, specifically \citet{1994A&AS..105..311F}, \citet{Danks:1994rt}, \citet{Allen:1995ys}, \citet{1996A&AS..117...93S}, \citet{Montes:1999vn}, and \citet{2000A&AS..146..217L}, as well as 80 bright stars we observed with UH2.2/SNIFS that are confirmed to be giants by {\it Hipparcos}. Many spectra have significantly higher resolution than our own observations. We convolve these data with a gaussian to match the resolution of our own sample to remove any resolution-dependency in our results.  To include sufficient SDSS photometry, we supplement our giant training set by including 200 giant stars with spectra from SDSS all with $r<16$ and proper motions consistent with zero. We require these SDSS spectra to have spectroscopic indices consistent with the rest of the giant training set. Because we select only SDSS stars with indices consistent with indices from spectra from the rest of the training stars, SDSS giant stars have no effect on our spectroscopic determination of luminosity class. Rather, these SDSS stars are added only for their photometry.

SDSS, 2MASS and WISE colors are available for much of our giant and dwarf training set; however, most lack $D51$ photometry, which covers the gravity-sensitive Mg Ib line at 5200\AA. Instead, we synthesize equivalent $g-D51$ colors from the spectra of our training set. We obtain the zero point for the synthesized colors of those stars in our sample which have both spectra and $g$ and $D51$ magnitudes. 

\subsection{Spectroscopic Determination of Luminosity Class}
\begin{figure}
\centering
\includegraphics[width=8cm]{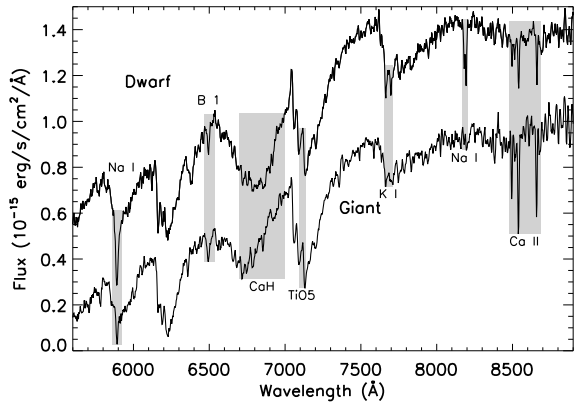}
\caption{SNIFS spectra of an M dwarf (top) and M giant (bottom) of similar $T_{eff}$ ($\simeq3600$~K) and magnitude ($K_P\simeq14$). Approximate regions for each of the six indices we use for giant/dwarf discrimination, as well as the TiO5 band, are marked in grey.  B 1 refers to a mix of atomic lines (Ba II, Fe I, Mn I, and Ti I) which overlap at the resolution of SNIFS ($\simeq 1300$).  The TiO5 molecular band is used as a probe of spectral type, although it is also sensitive to metallicity \citep{Lepine:2007fk}. Other atomic and molecular lines are generally much weaker in late type giant stars \citep{Reid:2005kx}. Indeed, the Na I (8172-8197\AA) and K I (7669-7705\AA) doublets are significantly weaker in the giant spectrum while they are both quite strong in the dwarf. \label{fig:giant_dwarf}}
\end{figure}

Our determination of luminosity class uses six different gravity-sensitive molecular or atomic indices (Table \ref{tab:indices} and Figure~\ref{fig:giant_dwarf}). Molecular and atomic indices are ratios of the average flux levels in a specified wavelength region to that of a pseudo-continuum region. Indices are useful for M dwarfs where the continuum is poorly defined. The values of most indices are a function of both gravity {\it and} temperature of the star. To remove this degeneracy we compare measured indices to the TiO5 spectral index. TiO5, as defined by \citet{Reid:1995lr}, is sensitive to spectral type and metallicity \citep{Woolf:2006uq, Lepine:2007fk} but it has minimal gravity dependance \citep{2008AJ....136..840J} (see Figure~\ref{fig:giant_dwarf}).

\begin{table*} 
\caption{Definitions of Spectroscopic Indices}
 \centering
\begin{tabular}{llll}
\hline \hline
Index Name & Band [\AA] & Continuum [\AA]& Source$^a$\\  \hline
Na I (a) & 5868-5918 & 6345-6355 & this work$^b$ \\
Ba II/Fe I/Mn I/Ti I& 6470-6530 &  6410-6420 & \citet{Torres-Dodgen:1993kx} \\
CaH2 & 6814-6846 & 7042-7046 & \citet{Reid:1995lr}\\
CaH3 & 6960-6990 & 7042-7046 & \citet{Reid:1995lr}\\
TiO5$^c$  & 7126-7135 & 7042-7046 & \citet{Reid:1995lr}\\ 
K I & 7669-7705 & 7677-7691, 7802-7825 & this work$^b$\\
Na I (b) & 8172-8197 & 8170-8173, 8232-8235& \citet{Schiavon:1997lr} \\
Ca II & 8484-8662 & 8250-8300, 8570-8600 & \citet{Cenarro:2001qy}\\
\hline
\end{tabular}
\tablecomments{Na I, Ba II/Fe I/Mn I/Ti I, K I , and Ca II are measured as equivalent widths, whereas CaH and TiO features are measured as band indices \citep{Reid:1995lr}.\\
$^a$ Papers where the wavelength definition we use is given. \\
$^b$ Wavelength ranges for Na I (b) and K I were determined from empirical analysis of the giant and dwarf training sets. \\
$^c$Because TiO5 has minimal gravity dependence, we measure other spectroscopic indices with respect to the TiO5 band strength.} 
\label{tab:indices}
\end{table*}

\begin{figure*}
\centering
\includegraphics[width=\textwidth]{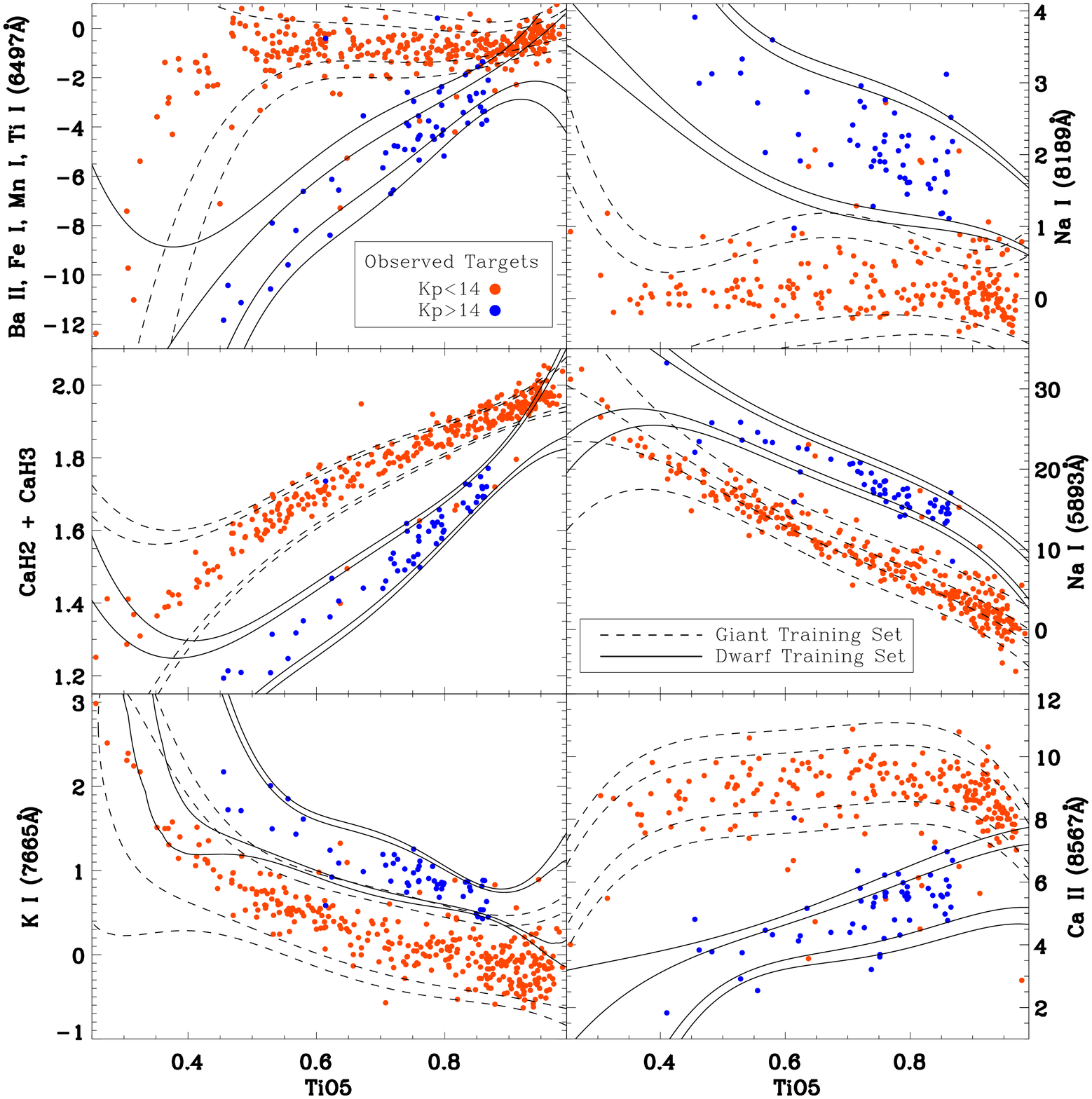}
\caption{Measured strengths of each gravity-sensitive spectral feature vs. the strength of the TiO5 band for {\it Kepler} late-type target stars with spectra from this program. Bright ($K_P<14$) targets are shown as red colored circles while faint ($K_P>14$) observed targets are shown as blue colored circles. The two-dimensional PDFs defined by our training set of giants (dashed line) and dwarfs (solid line) are overlaid. Contours of the PDF correspond to 68\%, and 90\%, intervals for the given training set. By using all spectral features, we positively identify each star with spectra as a giant or a dwarf with $>99\%$ certainty. \label{fig:giantdwarfgrid}}
\end{figure*}

We show spectra of giant and dwarf stars with similar effective temperatures in Figure~\ref{fig:giant_dwarf}, with the location of each feature labeled. As can be seen, most atomic lines are weaker in giants than in dwarfs. Indeed the Na I doublet (8172-8197\AA) and K I (7669-7705\AA) lines are quite shallow in giants while relatively deep in dwarfs \citep{Torres-Dodgen:1993kx,Schiavon:1997lr,Reid:2005kx}. Molecular lines provide additional luminosity-dependent spectral signatures. Metal hydride bands, such as the CaH bands defined by \citet{Reid:1995lr} and \citet{Lepine:2007fk} have been used for luminosity classification, although they are less useful for stars earlier than K7. The calcium triplet ($8484-8662$\AA) is a useful indicator of gravity \citep[e.g.][]{Cenarro:2001fk, Kraus:2009lr}, especially for M stars which emit comparatively more at red wavelengths. Giant and dwarf training sets overlaid on \kepler target star indices are shown in Figure~\ref{fig:giantdwarfgrid}. 

\subsection{Photometric Determination of Luminosity Class}\label{sec:giant_phot}
\begin{figure*}
\centering
\includegraphics[width=\textwidth]{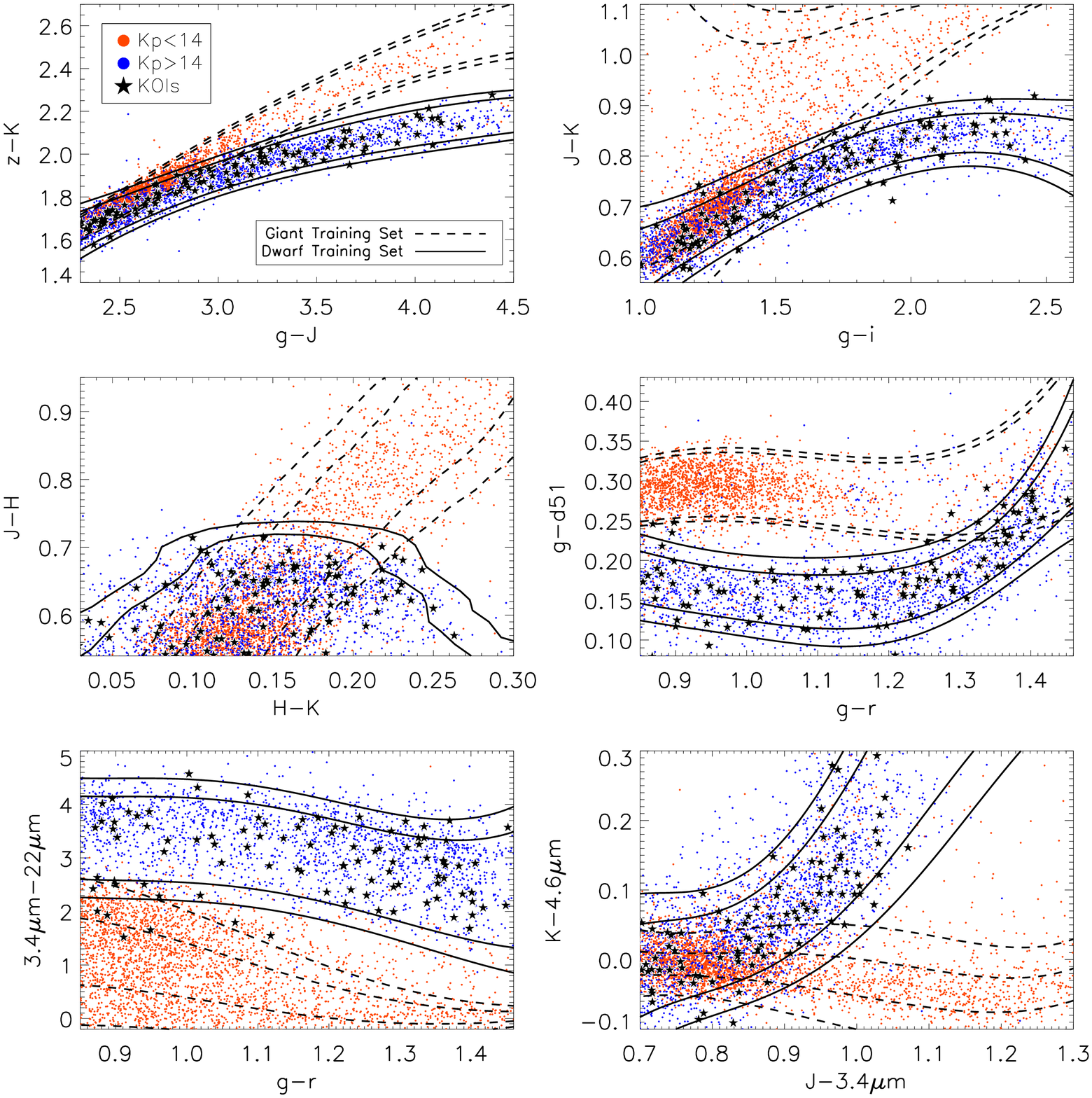}
\caption{Similar to Figure~\ref{fig:giantdwarfgrid} except using gravity-sensitive color-color relations. Each dot corresponds to a bright (red) or faint (blue) late-type {\it Kepler} target star. Contours are shown for the two training sets, corresponding to 68 and 90\% PDF intervals. We apply this cut to \kepler target stars with $J-H>0.52$ or $K_P-J>2.0$, although only a subsample of this set is shown for clarity. Most stars fall well inside either the dwarf or giant sequence, however, even when all color relations are used, $\simeq3\%$ of the sample still have an ambiguous luminosity class assignments. Most of these stars lack photometry in one or more band.  \label{fig:giantdwarfphotometry}}
\end{figure*}

We can use the available photometry to determine the luminosity class of a much larger sample of \kepler stars lacking spectra. \citet{Brown:2011fj} primarily use $g-D51$ vs. $g-r$ and $J-K$ vs. $g-i$ colors to separate \kepler late-type giants from dwarfs. Both giants and the coolest dwarfs in the sample have relatively weak Mg Ib lines, creating overlap between the dwarf and giant training sets at red $g-r$. A similar effect happens with $J-K$. Near-infrared photometry ($JHK$) has long been used to separate giants and dwarfs at redder colors \citep{Bessell:1988qy}, in part due to strong CO and weak Na I and Ca I absorption in giant stars. But for K and early M stars with $J-H<0.7$ and $H-K<0.2$, the giant and dwarf sequences overlap, creating a sizable region of ambiguity. At mid-infrared wavelengths, most giant stars have warm dust emission, leading to significantly redder colors in the WISE bandpasses.  Other relations can be derived from an examination of our giant and dwarf training sets. $z-K$ vs. $g-J$ follows a similar distribution to that of $J-K$ vs. $g-i$, but the giant and dwarf samples bifurcate at $g-J\simeq3.0$, which makes this color useful for isolating the reddest giants. Giant and dwarf training sets overlaid on \kepler target star colors are shown in Figure~\ref{fig:giantdwarfphotometry}.

\subsection{Application of training sets to the \kepler sample}\label{sec:apptokep} 
After each spectral index or color is measured or calculated for \kepler targets and both training sets, we identify stars as giants or dwarfs following the same technique as \citet{Gilbert:2006lr}. We begin by using the spectral index or color measurements of the training stars to produce a two-dimensional probability distribution function (PDF) for each index (or color). The PDFs are constructed by treating the strength of each index or color (henceforth $S$) as a Gaussian distributed variable with respect to $X$. For spectroscopic determination of luminosity class, $X$ is a parameter that primarily relates to the spectral type (although it may have some gravity dependence), while $S$ is a parameter that primarily relates to log~$g$. For the spectroscopic determination of luminosity class, $X$ is the TiO5 band and $S$ is one of our six gravity-sensitive indices (Na I, Ca II, Ba II/Fe I/Mn I/Ti I, K I, or CaH). For photometric determination of luminosity class, $X$ is defined as $g-J$, $g-i$, $J-H$, $g-r$, $3.4\mu$m - $22\mu$m, or $J-3.4\mu$m and $S$ is $z-K$, $J-K$, $H-K$, $g-D51$, $4.6\mu$m - $12\mu$m, or $K-4.6\mu$m, respectively. Values of $S$ are binned according to their corresponding $X$ value. Bins in $X$ are designed to contain an equal number of stars (20-25) in each bin, and because of this are not equally spaced in $X$. The mean ($\overline{S}$) and standard deviation ($\sigma_S$) of the distribution is computed in each bin. The two-dimensional PDF takes the form:
\begin{equation}\label{eqn:PDF}
PDF(X,S) = Cexp\left[\frac{-(S - \overline{S}(X))^2}{2(\sigma_{S}(X))^2}\right], 
\end{equation}
where $C$ is a normalization such that the entire PDF integrates to 1. PDFs for both giant and dwarf training sets overlaid on \kepler target star indices or colors are shown in Figures \ref{fig:giantdwarfgrid} and \ref{fig:giantdwarfphotometry} for the spectroscopic and photometric sets, respectively.

The likelihood that star $i$ is a dwarf for a given index $j$ is:
\begin{equation}
L_{i,j} = log_{10}\left( \frac{P_{dwarf}}{P_{giant}}\right),
\end{equation}
and the likelihood given all indices is:
\begin{equation}\label{eqn:weight}
\langle L_{i} \rangle = \frac{\sum_{j}w_j(X)L_{i,j}}{\sum_{j}w_j(X)}, 
\end{equation}
where $w_j$ is a weighting factor. Weights are calculated by determining the efficiency of a given feature at separating giants from dwarfs as a function of $X$. We take a random subsample (half the total sample) from each training set, and add Poisson noise to the spectra/colors consistent with our observations or given photometric errors. We then apply Equations \ref{eqn:PDF} - \ref{eqn:weight} to the subsamples using $w_j(X)=1$ for all $X, j$.  Values of $w_j$ are then set based on the fraction of dwarfs/giants correctly identified within a training set. $w_j(X) = 1$ if the feature/color identifies 100\% of the targets within a given $X$ bin correctly and $w_j(x) = 0$ if the feature/color identifies 50\% or less (i.e. no better than guessing) of the targets correctly. Weights are linearly interpolated (based on the fraction of stars correctly identified) between these two values.

\begin{figure}
\centering
\includegraphics[width=8cm]{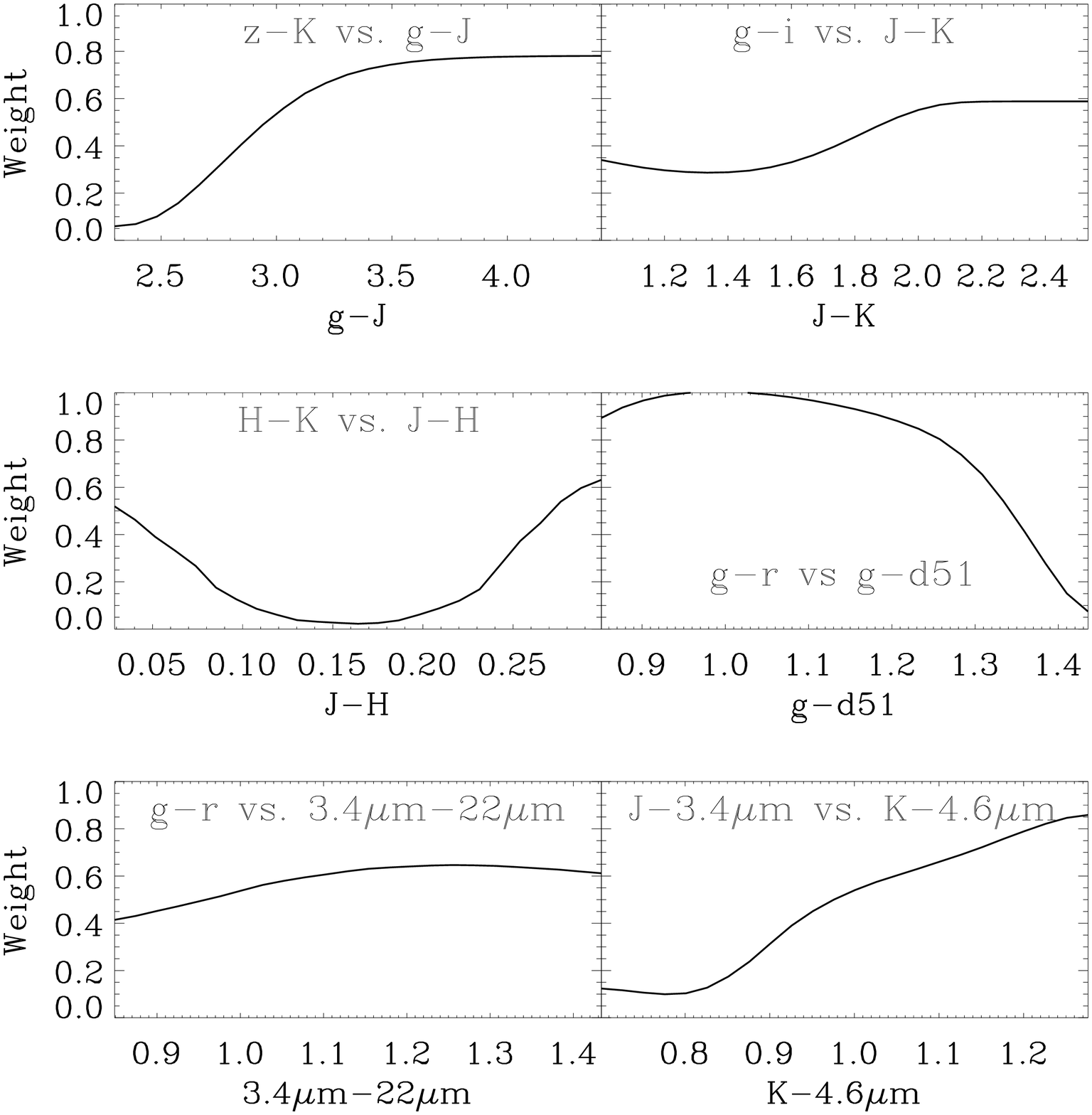}
\caption{Weights for each of the color relations used for our photometric determination of luminosity class. Weights are determined by applying the giant and dwarf PDFs derived from half of the training set to the other half (after adding Poisson noise to the data). Weights are set to 0 for all X outside of our training sets. Weights tend to be low in regions where the giant and dwarf training set PDFs overlap, or in regions where data is sparse.  \label{fig:weights}}
\end{figure}

Repeating the calculation of $L_{i}$ using $w_j=1$ for all $j$ does not change the classification of any stars with spectra (i.e. our results from spectra are essentially independent of our choice of weighting scheme). However, this is not the case for luminosity classes determined from color-color relations. The reason for this is the significant overlap between the PDFs of the color metrics for giant and dwarf training sets (e.g. $2.3 < g-J < 2.8$ and $1.6 < z-K < 1.9$, see Figure~\ref{fig:giantdwarfphotometry}). In overlapping regions, indices or colors will give similar probabilities for a star being a giant or a dwarf, making the metric less useful in giant/dwarf discrimination. This problem is solved by our weighting scheme, as regions where giant and dwarf training sets overlap tend to have lower weights. We show a plot of the weights for the color-color relations in Figure~\ref{fig:weights}. Weighting factors are set to 0 if any of the relevant indices/colors for a given star are missing or lie outside the range of our training sets. 

We identify all \kepler target stars with spectra as a giant or a dwarf with better than 99\% ($L_i>2.0$ or $L_i<-2.0$) confidence. The full list of determined luminosity classes for stars with spectra is given in Table \ref{tab:observations}. For the photometric sample, $\simeq97\%$ stars are placed into unambiguous giant or dwarf categories ($\langle L_{i} \rangle > 1.5$ for dwarfs or $\langle L_{i} \rangle < -1.5$ for giants). However, $\simeq 3\%$ of the sample are more ambiguous, most of which lack photometry in several bands. 

Since giant/dwarf assignments based on spectroscopy are very accurate, only binomial errors are considered for the spectroscopic sample. For uncertainty estimates from the photometric sample, we re-apply our likelihood calculations using 1000 different subsets of our training sets, adding random (Poisson) noise to the photometry, and then recalculating the giant fraction in each case. The variation in giant fraction is added in quadrature with binomial errors. This does not consider systematic errors (e.g. systematic photometric errors, discrepancies between training sets and \kepler target stars, etc). 

\subsection{Giant Star Fraction}
We find that, for the coolest \kepler stars ($K_P-J>2$), giant stars dominate the bright ($K_P<14$) \kepler target stars but are relatively rare among dim ($K_P>14$) targets. The fraction of giants is $96\pm1\%$ for bright stars, $7\pm3\%$ for dim stars, and $52\pm3\%$ for the combined set (based on our spectroscopy). Photometric assignments (considering $K_P-J>2$) give consistent giant fractions: $97\pm2\%$ for bright stars, $11\pm3\%$ for dim stars, and $55\pm3\%$ for all stars with $K_P-J>2$. The fractions in each brightness bin decrease somewhat when we apply a KIC log~$g>4.0$ cut. The giant fraction becomes $74\pm8\%$ for bright stars and $3\pm2\%$ for dim stars. The fraction of giants for all stars significantly decreases to $10\pm2\%$, due mainly to the large number of stars lacking any log~$g$ classification, most of which are giants and all of which are removed by this cut. 

\section{Planet occurrence}\label{sec:occurrence}
Following the work of H11, we calculate the planet occurrence, $f$, which is defined as the total number of planets, within a given range (in orbital period and radius) and considering all orbital inclinations, per star within a given range (in $T_{eff}$, log~$g$, and $K_P$). Planet occurrence will be somewhat higher than the fraction of stars with planets due to the presence of multi-planet systems, but if the rate of planet multiplicity is low, then these two quantities will be nearly identical.
\subsection{Nonparametric Estimation}\label{sec:simpleoccurrence}
We first calculate the planet occurrence following the nonparametric method of \citet{Gaidos:2012lr}. The total planet occurrence, $f$, is the sum of individual planet occurrences ($f_i$) over all $i$ planets that fall within a given range in orbital period and radius. The most probable occurrence of the $i$th {\it Kepler} detected planet in the population of $j$ {\it Kepler} target stars is:
\begin{equation}\label{eqn:freq}
f_i = \frac{1}{\displaystyle\sum\limits_{j=1}^{N} p_{i,j}d_{i,j}},
\end{equation}
where $d_{i,j} = 1$ if the S/N of a planet transit around the $j$th star is sufficient to detect the transit, and 0 otherwise, $p_{i,j}$ is the geometric probability of a transit, and $j$ is summed over all target stars that fall within a given range in $T_{eff}$, log~$g$, and $K_P$. We consider a planet detected ($d_{i,j} = 1$) if:
\begin{equation}\label{eqn:snr}
S/N = \frac{\delta}{\sigma_{CDPP}}\sqrt{\frac{N\tau}{30}} \ge 7,
\end{equation}
where $\delta$ is the transit depth, $N$ is the number of transits that occur over the observation interval, $\tau$ is the transit duration in minutes, and $\sigma_{CDPP}$ is the 30 minute combined differential photometric precision (CDPP) of {\it Kepler}. We use Quarter 1-2 30 minute CDPP values from {\it Kepler}. Our detection threshold $S/N = 7$ matches what is used by \citet{Borucki:2011uq} and \citet{Batalha:2012lr}. 

For small planets on nearly circular orbits,
\begin{equation}\label{eqn:prob}
p= 0.238P^{-2/3}M_*^{-1/3}R_*,
\end{equation}
where $P$ is the orbital period in days and $M_*$ and $R_*$ are the star's mass and radii in solar units. Values for $M_*$ and $R_*$ are computed by interpolating a grid of stellar radii/masses from the Dartmouth Stellar Evolution Database \cite[DSEP][]{Dotter:2008fk} at estimated values of $T_{eff}$, [Fe/H], and age. We use DSEP because radii and masses derived from their isochrones are in good agreement ($<0.03$ RMS deviation in radius) with current observations from interferometry \citep{Dotter:2008fk,Feiden:2011lr}.

For exoplanet hosts we use the metallicities given in M11, but for field stars metallicities are drawn from a random gaussian distribution of metallicities with $\overline{[Fe/H]}=-0.07$ and $\sigma_{[Fe/H]}=0.20$. This distribution is designed to be consistent with the distribution of M dwarfs in the solar neighborhood \citep{Johnson:2009fk, 2011A&A...530A.138C}. Ages are assigned randomly assuming a constant star formation rate (excluding ages $<100$~Myr). However, since M dwarfs do not change significantly while on the main sequence, our results are not changed when we fix all ages to 5~Gyr. The resulting stellar radii from the DSEP grid are used in conjunction with values of $R_p/R_*$ from \citet{Borucki:2011uq} to compute planetary radii. 

Estimates of $T_{eff}$ are inferred from our optical spectra. We compare our visible spectra to a grid of models of K- and M-dwarf spectra generated by the BT-SETTL version of PHOENIX \citep{Allard:2010fr}.  Details of the comparison, sub-grid interpolation,  and error calculations are described in L\'{e}pine et al. (in prep).  The grid of models spans $T_{eff}$ of 3000-5000~K in steps of 100~K, log~$g$ values of 0.0-5.0 in steps of 0.5~dex, and metallicities of [M/H] = -1.5, -1, -0.5, 0, +0.3, and +0.5.  $\alpha$/Fe is taken to be solar.  We report the $T_{eff}$ of the best-fit interpolated model, and the standard deviation of $T_{eff}$ among the set of interpolated models that are nearby in parameter space in Table~\ref{tab:observations}. 

\begin{figure}
\includegraphics[width=8.5cm]{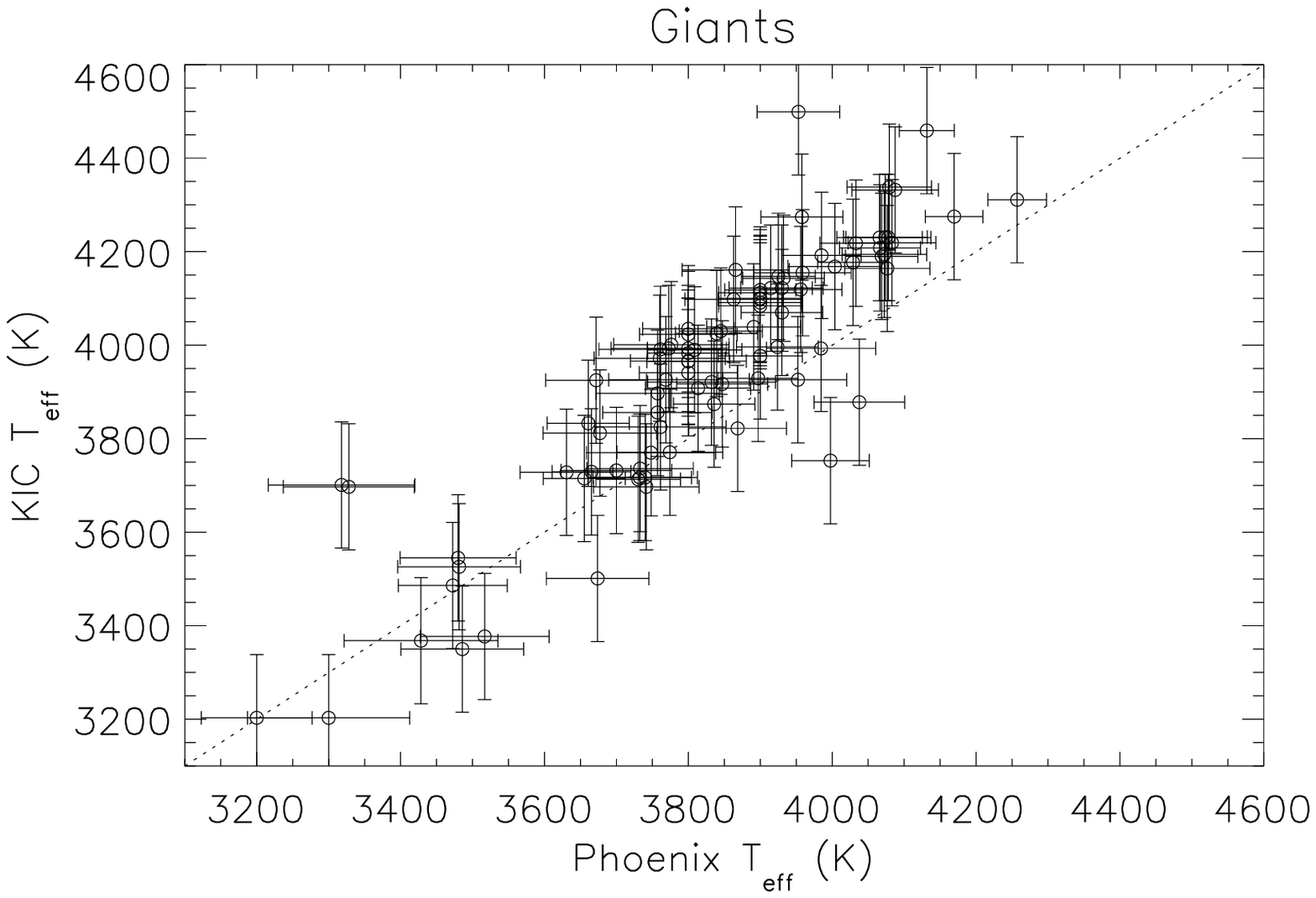}
\includegraphics[width=8.5cm]{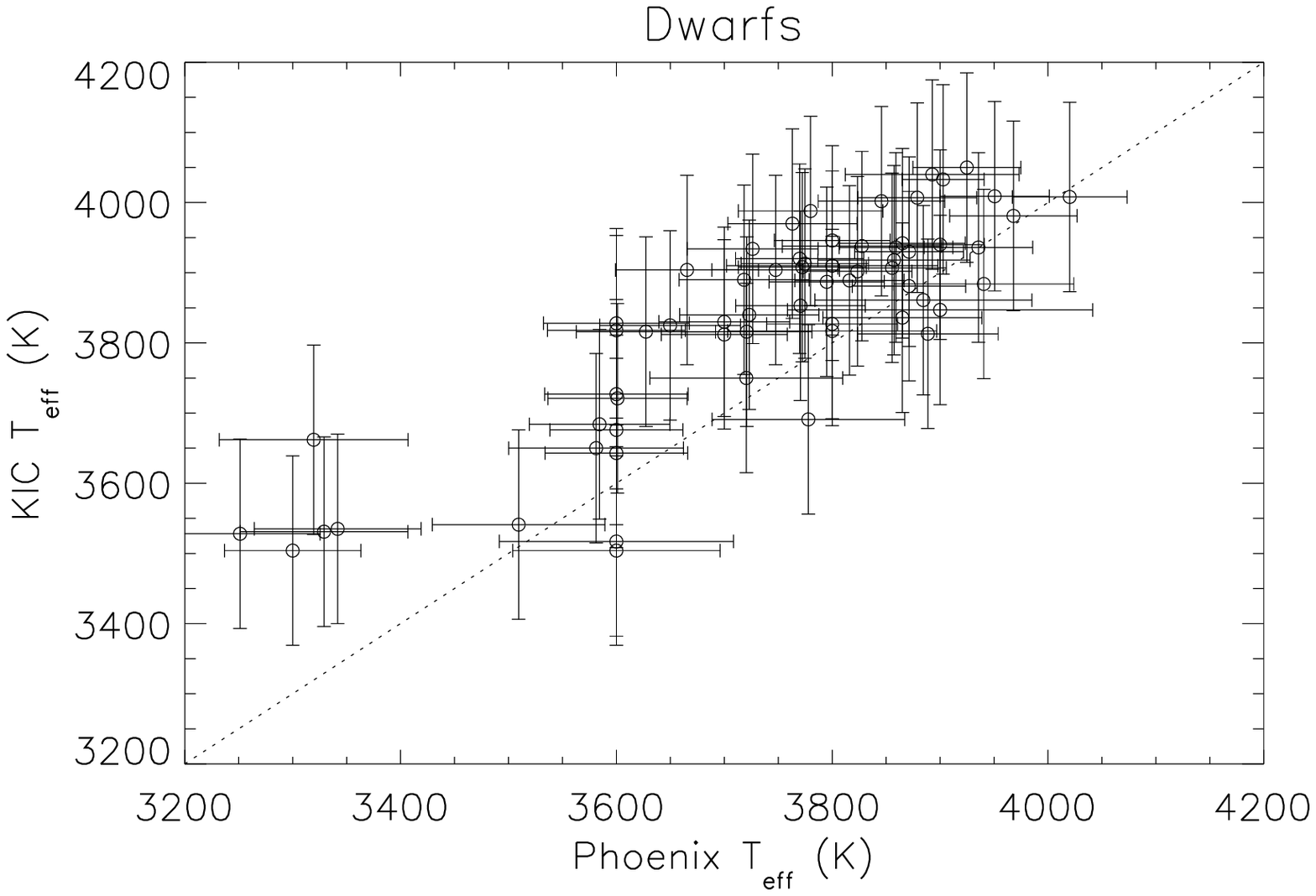}
\caption{Effective temperatures computed by fitting our spectra to models from the BT-SETTL version of PHOENIX \citep{Allard:2010fr} as a function of the KIC assigned effective temperature for giants (top) and dwarfs (bottom). The dotted line indicates equality. Errors are estimated as part of our model fitting procedure (errors on KIC temperatures are taken to be 135~K \citep{Brown:2011fj}. For both giants and dwarfs there is a clear 100-200~K offset between our spectroscopically determined temperatures and the KIC temperatures. This is most likely a consequence of the models used, as \citet{Castelli:2004lr} models used to fit KIC photometry to effective temperatures are unreliable below $4000$~K.} \label{fig:phoenixvskic}
\end{figure}

Our calculated values of $T_{eff}$ are shown in Figure~\ref{fig:phoenixvskic} vs. the temperature given in the KIC \citep{Brown:2011fj}. BT-SETTL temperatures are systematically lower than KIC temperatures by $110^{+15}_{-35}$~K for the dwarf stars, and $150^{+10}_{-35}$~K for the giant stars. Errors are calculated by bootstrap resampling. This is consistent with other determinations using the atmospheric models of \citet{Allard:2010fr}, including other determinations on \kepler KOI stars (M11). Our calculated temperatures are tightly correlated with KIC temperatures. When KIC temperatures are corrected for our observed offset, the standard deviation of the difference in calculated temperatures ($\sigma_{KIC-Phoenix}$) is 90~K, suggesting that the KIC temperatures for low-mass stars are {\it more} precise but are less accurate than suggested by \citet{Brown:2011fj}. For field stars with visible-wavelength spectra, we adopt our calculated $T_{eff}$ values, and for stars with exoplanet candidates we use the $T_{eff}$ from M11. For the remaining stars we adjust the KIC effective temperatures of \kepler stars downward randomly by $110^{+15}_{-35}$~K to keep the temperatures consistent with those of the KOI stars and those with spectra in our sample. This offset is randomized to account for errors in the systematic difference between temperatures calculated from our spectra and those listed in the KIC.

Following H11, we compute the planet occurrence with $2R_\earth<R_P<32_\earth$ and $P<50$ days around stars with $3400<T_{eff}<4100$ using Equations \ref{eqn:freq} - \ref{eqn:snr}. Again following H11, we exclude stars with $K_P>15$ where the accuracy of the planet candidate parameters are more questionable and the false positive rate is higher \citep{Morton:2011qy, Borucki:2011uq}. We calculate the standard deviation of the frequency using a Monte Carlo analysis. Stellar parameters are perturbed randomly (see above) accounting for errors from M11 on KOI metallicity and $T_{eff}$, and random errors from derived from $T_{eff}$ fits (see Figure~\ref{fig:phoenixvskic}) to our spectra. Other stars are given a random error of 90~K. We perturb transit parameters $R_P/R_*$ and period according to errors given by \citet{Borucki:2011uq}. Planetary radii are recalculated from perturbed values of $R_P/R_*$ and $R_*$. 

We remove planets from the KOI sample using the false positive probabilities from \citet{Morton:2011qy} (e.g., a planet candidate with a 5\% false positive probability is removed in 5\% of the simulations). We remove giant stars from the sample using the calculated photometric likelihoods (Section~\ref{sec:giant_phot}) for each star, such that a star with a 10\% likelihood of being a giant star will be removed from the sample in 10\% of the Monte Carlo (MC) simulations. This also applies to stars with detected planet candidates, causing the planet to be removed, i.e. we consider the planet detection to be a false positive if the star is a giant. The number of KOIs and target stars simulated varies somewhat for each Monte Carlo run, but there are typically $\simeq 14$ KOIs around $\simeq1300$ stars in a given simulation. 

We find that there are $0.37\pm0.08$ planets (with $2R_\earth < R_P < 32R_\earth$ and $P<50$ days) per star in the temperature range $3400 < T_{eff} < 4100$. For comparison we run an additional Monte Carlo simulation but only remove giant stars with KIC log~$g>4.0$ as in H11. This test yields a planet occurrence of $0.26\pm0.05$, slightly lower than when giant stars are properly removed. To test how our results depend on our choice of stellar radii model (DSEP) we also run two simulations using the Yonsei-Yale \citep{2004ApJS..155..667D} isochrones: one with giant stars removed as explained above and another removing just giants with KIC log~$g>4.0$. The runs using Yonsei-Yale are included because their models are commonly used to derive radii for {\it Kepler} targets \citep[e.g.][]{Batalha:2012lr}. However, radii and masses derived from DSEP are a far better match to observations of late-type stars \citep{Dotter:2008fk,Feiden:2011lr}, and planet occurrence calculated using the DSEP models should be considered more reliable. The resulting Monte Carlo distributions are shown in Figure~\ref{fig:planetfrequency}.

\subsection{Parametric Likelihood estimation}\label{sec:likelihood}

We also perform a parametric maximum likelihood estimation of the fraction of
stars with planets with radii $2R_{\oplus} < R < 32R_{\oplus}$ and
orbital period $P < 50$~d (see H11 for a similar analysis).  For discrete, binomial (detection or non-detection)
events, the likelihood is expressed as:
\begin{equation}
L = \prod_j^D \rho_j \times \prod_k^{ND} (1-\rho_k),
\end{equation}
where the first product is of detections, the second is of
non-detections, and $\rho_i$ is the probability that a planet with
properties in the appropriate ranges orbits the $i$th star and is detected by {\it Kepler} to transit.  For this formulation, we have assumed that $\rho \ll 1$.  We
adopt the specific power-law form $dN = CR_i^{-\alpha}P^{-\beta}d\ln R
\cdot d\ln P$ for the intrinsic distribution of planets.  If both
$\alpha$ and $\beta$ are $> 0$ then the normalization factor $C$ is
given by:
\begin{equation}
\label{eqn.normal}
C = \frac{f\alpha \beta}{\left(R_1^{-\alpha} - R_2^{-\alpha}\right)\left(P_1^{-\beta}-P_2^{-\beta}\right)},
\end{equation} 
where $f$ is the total planet occurrence.  We do not
model multi-planet systems; that level of analysis is not justified
given the large uncertainties in our parameters.

Following the usual procedure, we maximize the logarithm of $L$:
\begin{eqnarray}\nonumber
\ln L &=& \sum_j^D \left[\ln C - \alpha \ln R_j - \beta \ln P_j+ \ln D_j(R_j,P_j) \right]  \\
&&  + \sum_k^{ND} \ln \left[1-CF_k(\alpha,\beta)\right]
\end{eqnarray}
where $D_j(R_j,P_j)$ is the probability of detecting the $j$th planet
around its host star, including the geometric factor (note $D_j(R_j,P_j) = d_{j}p_{j}$, see Equation \ref{eqn:freq} and \ref{eqn:prob}), and
\begin{equation}
  F_k(\alpha,\beta) = \int_{R_1}^{R_2}\int_{P_1}^{P_2}R^{-\alpha}P^{-\beta}D_k(R,P)d\ln R \cdot d\ln P
\end{equation}
If the detection rate is low, then:
\begin{eqnarray}\nonumber
\ln L &\approx& \sum_j \left[\ln C - \alpha \ln R_j - \beta \ln R_j + \ln D_j(R_j,P_J) \right] \\
&&- C\sum_k^{ND} F_k(\alpha,\beta).
\end{eqnarray}
We then substitute Equation \ref{eqn.normal} for $C$.  Ignoring terms
that do not depend on $\alpha$, and thus do not affect its maximum likelihood value, we
find the following quantity must be maximized:
\begin{eqnarray}\nonumber
\label{eqn.6}
\ln L_{\alpha} &=& \sum_j \left[\ln \alpha - \ln\left(R_1^{-\alpha} - R_2^{-\alpha}\right) - \alpha \ln R_j \right] - \\
&&\frac{f\alpha \beta \sum_k^{ND}F_k(\alpha,\beta)}{\left(R_1^{-\alpha} - R_2^{-\alpha}\right)\left(P_1^{-\beta}-P_2^{-\beta}\right)}.  
\end{eqnarray}
Likewise,
\begin{eqnarray}\nonumber
\label{eqn.7}
\ln L_{\beta} &=& \sum_j \left[\ln \beta - \ln\left(P_1^{-\beta} - P_2^{-\beta}\right) - \beta \ln P_j \right] - \\
&&\frac{f\alpha \beta \sum_k^{ND}F_k(\alpha,\beta)}{\left(R_1^{-\alpha} - R_2^{-\alpha}\right)\left(P_1^{-\beta}-P_2^{-\beta}\right)}.  
\end{eqnarray}
The simultaneous solution for the planet occurrence is
found by maximizing the terms that depend on $f$ and is simply
\begin{equation}
\label{eqn.8}
f = \frac{N_p\left(R_1^{-\alpha} - R_2^{-\alpha}\right)\left(P_1^{-\beta}-P_2^{-\beta}\right)}{\alpha \beta \sum_k^{ND}F_k(\alpha,\beta)},
\end{equation}
where $N_p$ is the number of detected planets.  Equation \ref{eqn.8}
immediately suggests a reduction in the last terms of Equations
\ref{eqn.6} and \ref{eqn.7} to $N_p$, which is independent of $\alpha$
and $\beta$ and can be ignored.  

\begin{figure}
\mbox{
\includegraphics[width=8.5cm]{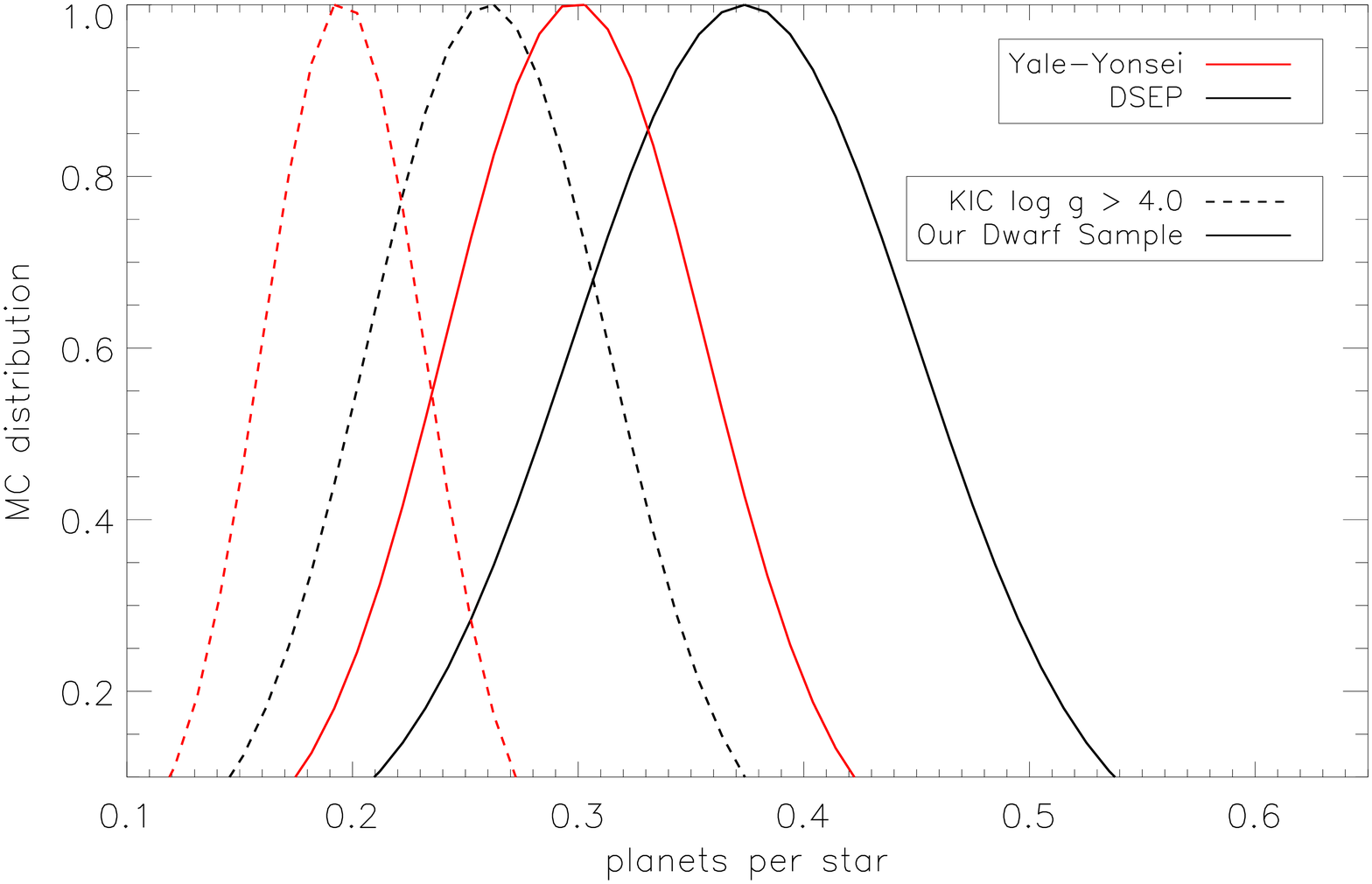}}
\quad
\includegraphics[width=8.5cm]{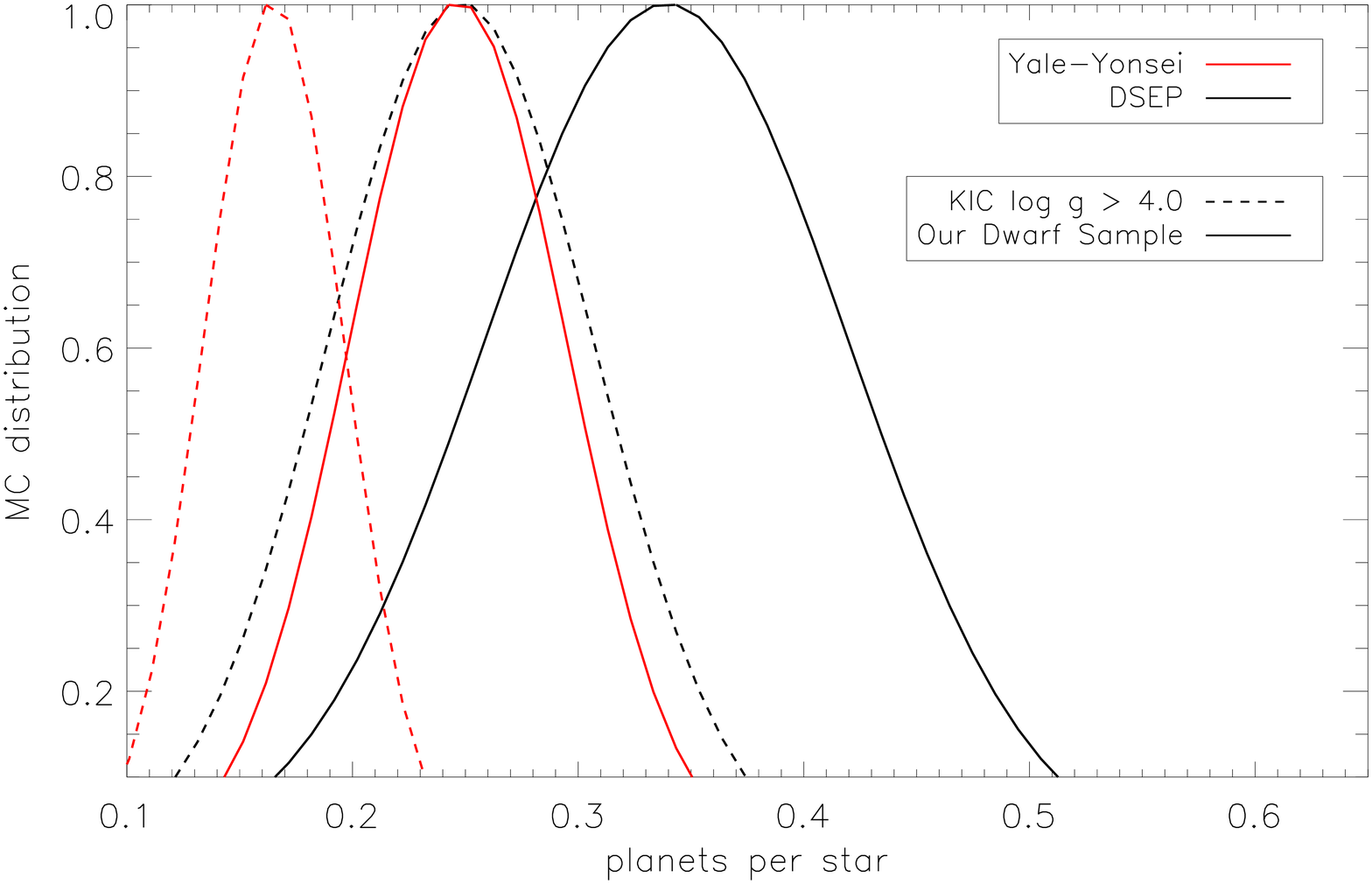} 
\caption{Planet occurrence with giant stars removed (solid line) or using KIC log~$g>4.0$, using isochrones from DSEP (black) or from Yonsei-Yale (red) calculated by Monte Carlo analysis. The top plot is calculated using nonparametric MC estimate, and the bottom uses a parametric MC estimate. For both plots, we consider planets with radii $2R_\earth<R_P<32_\earth$ and periods $P<50$ days, and stars with effective temperatures $3400<T_{eff}<4100$. A full description of our analysis is given in Section~\ref{sec:occurrence}. \label{fig:planetfrequency}}
\end{figure}

Because there are too few systems in our sample to get a robust estimate of $\beta$, we fix $\beta = 0$ with a cut-off at $P_1 = 1$~d,
consistent with the findings of previous analyses
\citep[][H11]{Cumming:2008bh,Wolfgang:2011uq}.  Equation \ref{eqn.8} becomes:
\begin{equation}
f = \frac{N_p\left(R_1^{-\alpha} - R_2^{-\alpha}\right)\ln (P_2/P_1)}{\alpha \sum_k^{ND}F_k(\alpha,\beta=0)}.
\end{equation}
Artificial Monte Carlo data sets suggest that $f$ is robustly recovered, but that recovered values of $\alpha$ are biased
downwards. Using the cool KOIs defined here, stellar parameters derived as explained above, and Monte Carlo data sets generated by sampling with replacement, we find that $f = 0.34 \pm 0.08$, consistent with our nonparametric calculation. As before, we repeat our Monte Carlo simulation but only removing giant stars with KIC log~$g>4$, and another run using the Yonsei-Yale evolutionary tracks \citep{2004ApJS..155..667D} instead of those of DSEP. The resulting Monte Carlo distributions are shown in Figure~\ref{fig:planetfrequency}. \\

\section{Planet-host Metallicities}\label{sec:metallicity}
SL11 use $g-r$ vs. $J-H$ colors to conclude that late-type ($J-H\simeq0.62$) exoplanet hosts are redder and more metal-rich than stars without transiting planets. Because giant stars have bluer $g-r$ colors at a given $J-H$ color \citep{Bessell:1988qy, Gilbert:2006lr}, a significant number of giant star interlopers in their sample will cause field stars to appear metal poor. Giant stars have stellar radii 10-100 times larger than dwarfs, significantly reducing the depth in a light curve for a given transiting planet, making it much less likely that they will appear as KOIs (with the exception of false positives). 

We can test their findings by creating a ``pure'' dwarf sample, and comparing its color distribution to that of the KOI sample. Our $K_P-J>2$ spectroscopic sample is systematically redder in $J-H$ than the $0.56<J-H<0.66$ bin used in SL11, preventing us from making a direct comparison. Instead, we construct samples of giants and dwarfs in the $J-H\simeq0.62$ bin based on our photometric determination of luminosity class. For both the dwarf and giant samples, we select \kepler target stars with photometry in all bands used in our photometric assignment of luminosity class ($J, H, K, D51, g, r, $ and all four WISE bands). We then select stars with a $> 90\%$ likelihood of being dwarfs based on our analysis in Section~\ref{sec:giant_phot}. The resulting dwarf sample is $\simeq 2500$ stars. This sample may still contain giants. We add Poisson noise to the photometry of both the training sets and the \kepler $0.56<J-H<0.66$ target star sample, and take random subsamples of both training sets. We then reapply these subsamples to the modified photometry of the \kepler sample. We repeat this process 1000 times. By analyzing the number of giant stars in each of these new samples we find that our dwarf sample is $<1\%$ giant stars at 95\% confidence, ignoring possible systematic errors.

\begin{figure}
\centering
\includegraphics[width=8cm]{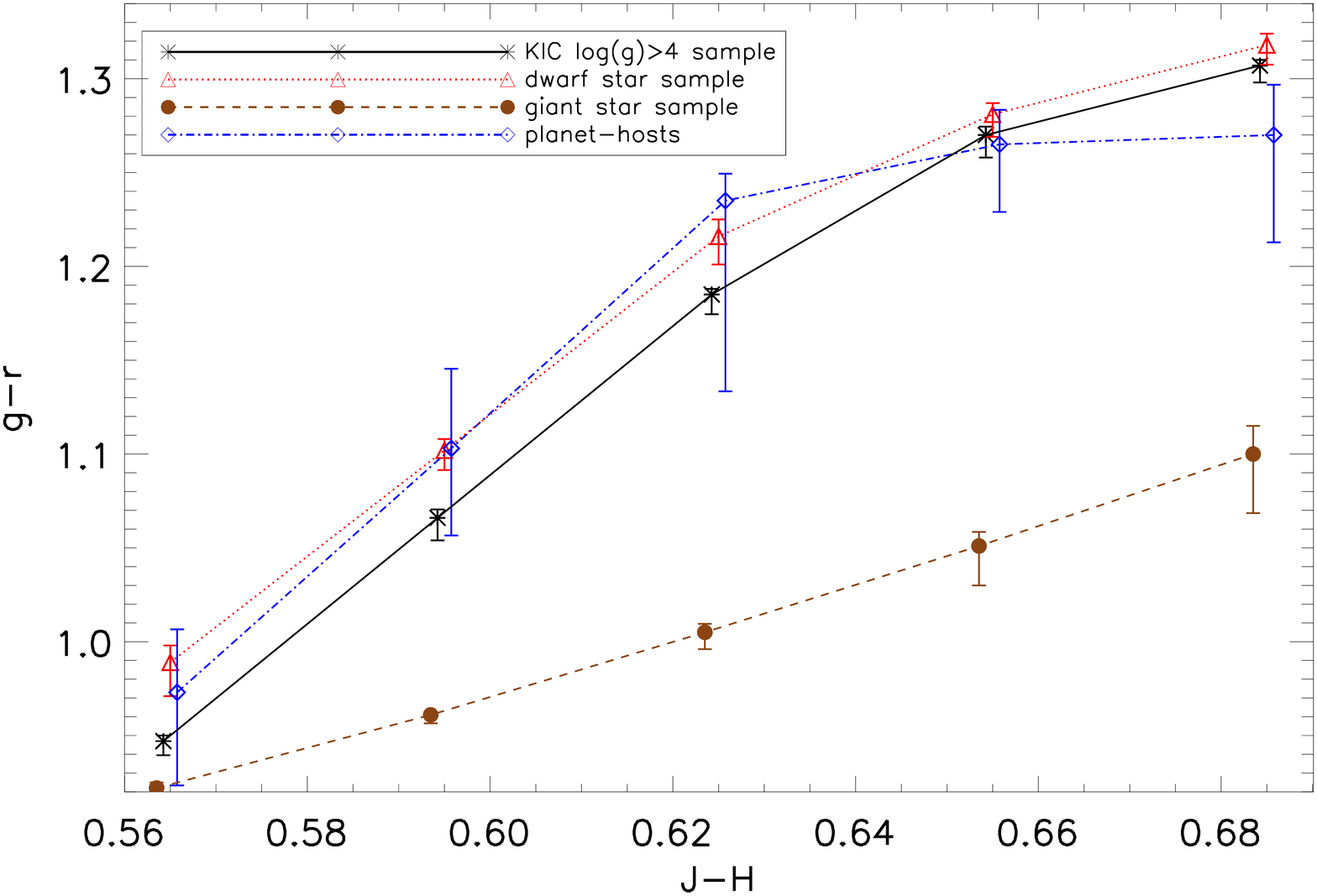}
\caption{Median $g-r$ colors as a function of $J-H$ colors for \kepler target stars with: Earth to Neptune sized planet candidates (dotted/dashed line, diamonds), KIC log~$g>4.0$ (solid line, asterisks), $>90\%$ likelihood of being dwarfs based on their colors (dotted line, triangles), $>90\%$ likelihood of being giants (dashed line, circles). The $1\sigma$ errors are calculated for the median in each bin by bootstrap resampling. Bins for all data sets are the same, but each point is offset slightly from the bin center for clarity. There is a statistically significant offset between the KIC log~$g>4.0$ sample and the planet hosts when we consider stars with $0.58<J-H<0.66$, however, this offset is no longer present when misidentified giant stars are removed from the sample. Indeed, our dwarf control sample closely tracks the colors of the planet-hosting stellar population. \label{fig:planetmetal}}
\end{figure}

We use this dwarf sample, following the method of SL11, to compare the $g-r$ colors at a given $J-H$ (a proxy of effective temperature) of the exoplanet host stars with our dwarf sample. Figure~\ref{fig:planetmetal} shows $g-r$ colors as a function of $J-H$ colors for the dwarf, giant, planet-host, and KIC log~$g>4.0$ sample. We find no significant difference in color between the KOI stars and our dwarf sample. Unlike the KIC log~$g>4.0$ sample, {\it the locus of our photometrically selected dwarf sample is consistent with the locus of the KOI sample at $J-H\simeq0.62$}. For stars with $K_P-J>2.0$ we find an offset in $g-r$ color of only $-0.05\pm0.03$ between the spectroscopically confirmed dwarfs and late-type KOI stars hosting Earth-to-Neptune sized planets. When we use our photometric sample of dwarfs in the $J-H\simeq0.62$ bin we find an offset of $0.01\pm0.02$ and we can rule out the offset of 0.08 seen by SL11 with $>99.7\%$ certainty. Our photometric selection may remove some metal-poor dwarfs. However, even when we include stars $\ge60\%$ likelihood of being dwarfs, which will necessarily increase the number of interloping giants, the offset is still only $0.03\pm0.02$ (consistent with zero offset). 

\begin{figure*}
\centering
\includegraphics[width=\textwidth]{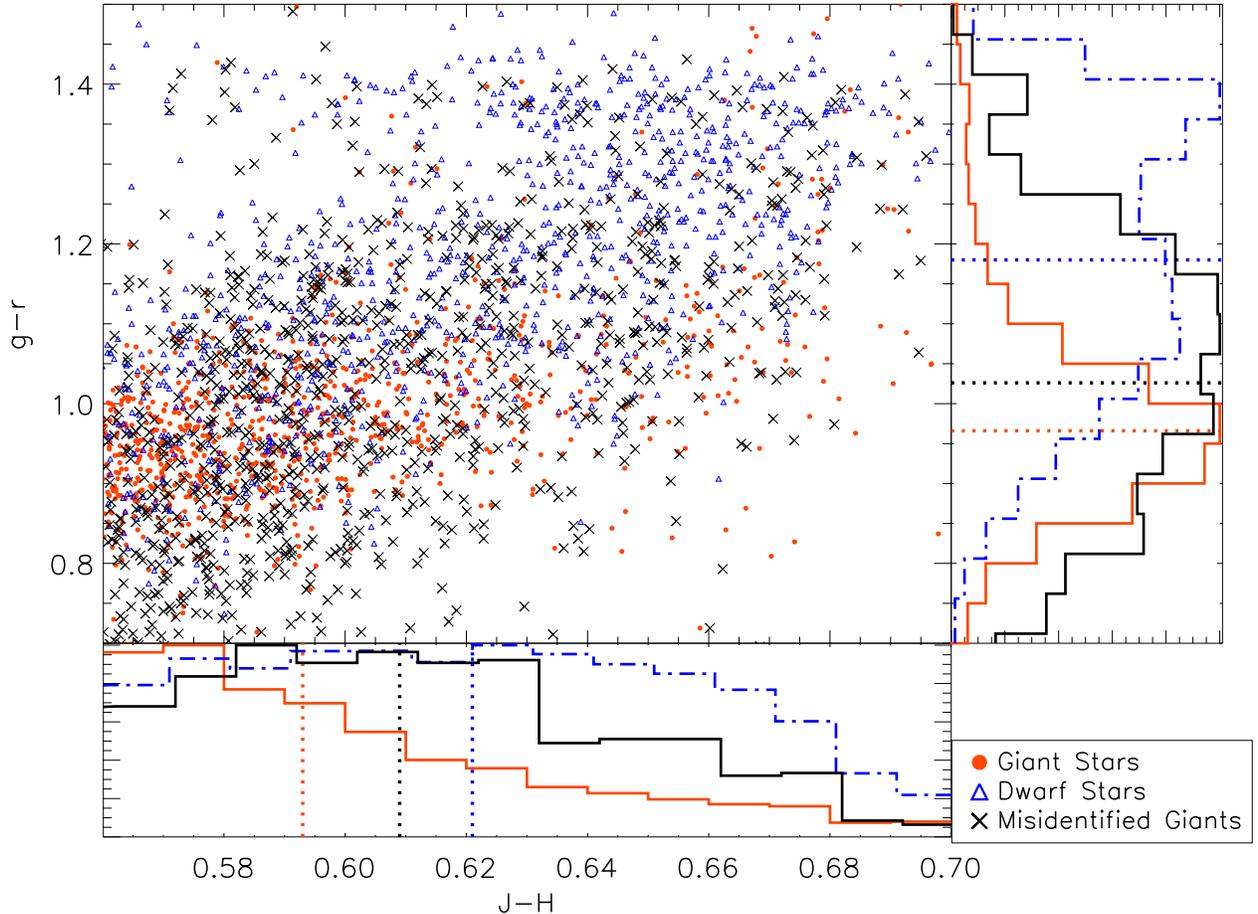}
\caption{Scatter plot of a sample of {\it Kepler} dwarf stars (blue triangles), giant stars (red circles), and giant stars labeled as dwarfs (KIC log~$g>4.0$) by the KIC (black crosses) in $g-r$ vs $J-H$ space. An equal number of data points are shown from each subset (giants/dwarfs/misidentified giants) to highlight the {\it relative} distributions. The histograms on the bottom and right side show the 1-D distribution in each color (coloring matches the center plot). Histograms are normalized to a peak value of 1 and the median of each histogram is marked with a dotted line (of the corresponding color). Although giant stars cover a range of $J-H$ colors, those that were mislabeled as dwarfs are more concentrated around $J-H\simeq0.61$. The distribution of misidentified giants is bluer that the dwarf distribution. Thus if the misidentified giant stars are not properly removed the dwarf sample will appear bluer (more metal poor) than the KOI distribution (which contains almost no misidentified giants). \label{fig:jhphot}}
\end{figure*}

In spite of the low giant fraction for dim \kepler target stars, it is not sufficient to simply repeat the SL11 analysis exclusively for stars with $K_P>14$. Since SL11 only examine stars with KIC log~$g>4.0$, it is far more important to investigate the $g-r$ distribution of {\it misidentified} giants in the $0.56<J-H<0.66$ color range (i.e. giant stars that were assigned log~$g>4$ in the KIC). In fact the fraction of misidentified dim giant stars in their $J-H\simeq0.62$ bin is higher ($12\%$), than it is for the $K_P-J>2$ star sample. We show why this is the case in Figure~\ref{fig:jhphot}, which shows the distribution of giants, dwarfs, and misidentified giants in $J-H$ vs. $g-r$ space. Misidentified giants are more concentrated at $0.58<J-H<0.63$. Further, the misidentified giants in this $J-H$ range are much more blue than the dwarfs in the same range. Thus by selecting a color bin centered on $J-H=0.62$, SL11 are over-selecting giant stars, even after applying a KIC log~$g>4$ cut ($\simeq15\%$ of this sample are giant stars). This concentration of misidentified giants is the most likely explanation for the color offset seen by SL11, and also explains why the same $g-r$ offset is not seen at redder $J-H$ colors (see Figure~\ref{fig:planetmetal}). 

\section{Discussion}
We use visible-wavelength spectra to determine the properties of a subset of late-type \kepler target stars. We separate giants from dwarfs by comparing our spectra to those of stars with known luminosity class, and determine effective temperatures by comparing with PHOENIX model spectra. We extend our results to a larger collection of \kepler stars using photometry from the KIC, 2MASS, and WISE catalogs. We apply our luminosity class determinations to refine estimates of the planet occurrence around stars with $3400<T_{eff}<4100$, and compare the colors -- and hence metallicities of stars with and without detected Earth and Neptune sized planets. We draw four major conclusions:

1. Among stars redder than $K_P-J=2$ ($\simeq$ K5 and later), bright ($K_P<14$) stars are predominantly ($96\pm1\%$) giants, while dim stars ($K_P>14$) are predominantly ($93\pm3\%$) dwarfs. These fractions improve somewhat when we consider stars with KIC determined log~$g > 4.0$ ($74\pm8\%$ and $97\pm2\%$ respectively).  Overall, $52\pm3\%$ of \kepler stars with $K_P-J>2$ are giants. However, only $10\pm2\%$ of said stars with KIC log~$g>4.0$ are giants, a consequence of the large number of late-type stars lacking any temperature or log~$g$ values in the KIC.

2. KIC effective temperatures, based on the models of \citet{Castelli:2004lr} and $griz$ and $JHK$ photometry, are systematically higher by $110^{+15}_{-35}$~K compared to those derived from our own spectra and PHOENIX BT-SETTL atmosphere models \citep{Allard:2010fr}.

3. Adopting the temperature scale from BT-SETTL and radii/masses from the Dartmouth Stellar Evolution Database \citep{Dotter:2008fk} and removing stars we identify as giants based on nonparametric and parametric Monte Carlo calculations we find a planet occurrence rate of $\simeq 0.36\pm0.08$ for planets with radii $2R_\earth < R_P < 32R_\earth$ and periods $1<P<50$~days per star in the temperature range $3400 < T_{eff} < 4100$. Using the KIC determined luminosity classes leads to a somewhat lower planet occurrence of $0.26\pm0.05$. 

4. The $g-r$ colors of exoplanet host stars at $J-H\simeq0.62$ are consistent with an unbiased sample of \kepler dwarf stars, ruling out any large difference between hosts of Earth-to-Neptune sized planets and those without any detected planets. 

Surprisingly, there are hundreds of stars in our photometric sample that could have been easily identified as giants with KIC photometry, but were assigned log~$g>4$.  The KIC primarily uses $g-D51$ vs $g-r$ colors to identify giants, and many late-type stars with KIC log~$g>4.0$ have $g-D51$ vs $g-r$ colors consistent with giants (and inconsistent with dwarfs). 

Our calculated giant fraction is consistent with other independent measurements. \citet{Gaidos:2012lr} compare radial velocity data from M2K \citep{Apps:2010zr,Fischer:2012lr} to \kepler results and note that the completeness of the coolest \kepler target stars may be quite low ($\simeq 50\%$), much of which could be explained by an underestimate of the frequency of giant stars. Additionally, \citet{Ciardi:2011lr} find that bright \kepler M stars are``predominantly giants, regardless of the KIC classification'' based on $JHK$ photometry alone. Our giant fraction is also consistent with the current understanding of Galactic structure: based on a simulation from TRILEGAL \citep{2005A&A...436..895G}, $\simeq92$ of stars near the center of the {\it Kepler} field with $K_P<14$ and $K_P-J > 2.0$ are giants.

Interestingly, we find two KOIs with colors consistent with giant stars. KOI 667 and KOI 977 both fall within our giant training set in multiple color relations, and outside our dwarf training set. M11 identify KOI 977 as a giant, and they also note that KOI 667 consisted of 5 objects within $6\arcsec$ which may be contaminating 2MASS or WISE photometry. One of these objects could be an eclipsing binary, diluted by the other stars. KOI 667 also has a relatively high (10\%) false positive probability based on Galactic structure models \citep{Morton:2011qy}.

Our values of $T_{eff}$ are consistent with results reported elsewhere also using BT-SETTL, including observations of the late-type KOIs with near-infrared spectra M11. These authors find a similar systematic offset of $123^{+24}_{-32}$~K between their temperatures and KIC assigned temperatures. KIC temperatures are based on the models of \citet{Castelli:2004lr} and the evolutionary tracks of \citet{2000A&AS..141..371G}, which, although reliable for solar-mass stars, are untrustworthy for stars with $T_{eff}<3750$~K \citep{Brown:2011fj}.

Our planet occurrence estimate is slightly higher than that of H11, who, using results from {\it Kepler}, find a planet occurrence rate of $0.30\pm 0.08$ for stars with $3600<T_{eff}<4100$. The difference is primarily due to reliance on luminosity class determinations by \citet{Brown:2011fj}, which we find to be inaccurate. However, the difference is within $1\sigma$. For both our work and that of H11, errors are dominated by the low number of late-type stars (and therefore planets around them) in the \kepler field and very high random ($\sim35\%$) errors in stellar radii. 

In addition to random errors (e.g. stellar radii and $R_p/R_*$) that are included in our Monte Carlo simulation, there may be large systematic uncertainties in atmosphere models and evolutionary tracks, which can change the resulting frequency. When we use the Yonsei-Yale isochrones, it decreases our planet occurrence by $\simeq0.08$. Interestingly, this difference is  similar in size to the random errors in our Monte Carlo analysis ($\simeq0.08$), and the difference between proper giant removal and using KIC log~$g>4.0$ ($\simeq0.10$). This suggests that giant star removal, improved stellar characterization of the dwarf stars, and use of reliable stellar models of late-type stars are of roughly equal importance in characterizing the planet occurrence around very cool stars.

The lack of a strong correlation between host-star metallicity and the presence of Earth-to-Neptune sized planets is consistent with what is found for solar-type stars, e.g. \citet{Mayor:2011fj}. This also matches the findings of M11, who determine that among the late-type \kepler exoplanet hosts in our sample the median [M/H] is $-0.11\pm0.02$. This distribution is consistent with stars in the solar neighborhood \citep[$-0.05$ to $-0.15$,][]{Johnson:2009fk, Schlaufman:2010qy, 2011A&A...530A.138C}. A metallicity difference could only be present if \kepler target stars are significantly more metal poor than stars in the solar neighborhood. As explained in \citet{Gaidos:2012lr}, {\it Kepler} late K and M stars are $<250$~pc from the Sun, and $\lesssim60$~pc above the galactic plane. Most of the stars will be in the thin disk, and have metallicities similar to that of the solar neighborhood.

Our analysis of the $g-r$ colors of planet hosts contradicts the results of SL11, who find a $4\sigma$ difference between $g-r$ colors of late-type exoplanet hosts and stars with no exoplanets present. Their result is most likely an artifact of the large number of stars which were misclassified as dwarfs in the KIC. SL11 state that their result can be reproduced if their sample of KIC log~$g>4$ stars is between 10\% and 30\% giants, which they calculate by adding stars with KIC log~$g<4$ stars (test giant stars) into their control sample, and measuring the resulting $g-r$ color offset. We find that the giant fraction is above $10\%$ for this color range. Further, if the KIC log~$g>4$ sample that SL11 used was significantly contaminated with giants, the sample will have bluer colors than a true dwarf sample. Adding test giants (to measure the resulting color change) to an already giant-star contaminated sample will create smaller changes in the overall color of a sample than if the sample had contained only dwarf stars. Thus more test giant stars will be required to produce a given color offset, creating an artificially high estimate for the level of giant contamination required to produce the observed color difference.

Although the $g-r$ colors of exoplanet hosts in our sample are consistent with our dwarf sample, we cannot rule out small offsets ($\lesssim0.05$) in $g-r$ color. It is possible that any metallicity effect is sufficiently small that it is diluted to non-detection by the large number of undetected exoplanets in the dwarf sample. As \kepler continues to discover planets of smaller radii and at larger orbital periods, the answer may become more clear.

\acknowledgments
This work was supported by NSF grant AST-0908419, NASA grants NNX10AI90G and NNX11AC33G (Origins of Solar Systems) to EG, as well as NSF grants AST 06-07757 and AST 09-08419 to SL. We thank the anonymous reviewer, whose thoughtful, quick, and thorough comments helped make this manuscript significantly better.

SNIFS on the UH 2.2-m telescope is part of the Nearby Supernova Factory project, a scientific collaboration among the Centre de Recherche Astronomique de Lyon, Institut de Physique NuclŽaire de Lyon, Laboratoire de Physique NuclŽaire et des Hautes Energies, Lawrence Berkeley National Laboratory, Yale University, University of Bonn, Max Planck Institute for Astrophysics, Tsinghua Center for Astrophysics, and the Centre de Physique des Particules de Marseille.

Some of the data presented in this paper were obtained from the Multimission Archive at the Space Telescope Science Institute (MAST). STScI is operated by the Association of Universities for Research in Astronomy, Inc., under NASA contract NAS5-26555. Support for MAST for non-HST data is provided by the NASA Office of Space Science via grant NNX09AF08G and by other grants and contracts. 

Funding for the SDSS and SDSS-II has been provided by the Alfred P. Sloan Foundation, the Participating Institutions, the National Science Foundation, the U.S. Department of Energy, the National Aeronautics and Space Administration, the Japanese Monbukagakusho, the Max Planck Society, and the Higher Education Funding Council for England. The SDSS Web Site is http://www.sdss.org/.

This publication makes use of data products from the Wide-field Infrared Survey Explorer, which is a joint project of the University of California, Los Angeles, and the Jet Propulsion Laboratory/California Institute of Technology, funded by the National Aeronautics and Space Administration.

{\it Facilities:} \facility{{\it Kepler}}, \facility{MKO}, \facility{UH2.2m}, \facility{KPNO}, \facility{MDM}

\bibliography{/Users/andrewmann/dropbox/fullbiblio.bib}

\clearpage

\end{document}